\documentclass[conference]{IEEEtran}
\IEEEoverridecommandlockouts
\usepackage{cite}
\usepackage{amsmath,amssymb,amsfonts}
\usepackage{algorithmic}
\usepackage{graphicx}
\usepackage{textcomp}
\usepackage{xcolor}
\usepackage{cite}
\usepackage{url}
\usepackage{booktabs}
\def\BibTeX{{\rm B\kern-.05em{\sc i\kern-.025em b}\kern-.08em
    T\kern-.1667em\lower.7ex\hbox{E}\kern-.125emX}}
\begin{document}

\title{Emo-DPO: Controllable Emotional Speech Synthesis through Direct Preference Optimization}

\author{\IEEEauthorblockN{1\textsuperscript{st} Xiaoxue Gao}
\IEEEauthorblockA{
\textit{Agency for Science, Technology, and Research}\\
Gao\textunderscore Xiaoxue@i2r.a-star.edu.sg}
\and
\IEEEauthorblockN{2\textsuperscript{nd} Chen Zhang}
\IEEEauthorblockA{
\textit{National University of Singapore}\\
chen\textunderscore zhang@u.nus.edu}
\and
\IEEEauthorblockN{3\textsuperscript{rd} Yiming Chen}
\IEEEauthorblockA{
\textit{National University of Singapore}\\
yiming.chen@u.nus.edu}
\and
\IEEEauthorblockN{4\textsuperscript{th} Huayun Zhang}
\IEEEauthorblockA{
\textit{Agency for Science, Technology, and Research}\\
Zhang\textunderscore Huayun@i2r.a-star.edu.sg}
\and
\IEEEauthorblockN{5\textsuperscript{th} Nancy F. Chen}
\IEEEauthorblockA{
\textit{Agency for Science, Technology, and Research}\\
nfychen@i2r.a-star.edu.sg}
% \textit{Institute for Infocomm Research, Agency for Science, Technology, and Research}\\
}
\maketitle
\vspace{-0.5cm}
\begin{abstract}
Current emotional text-to-speech (TTS) models predominantly conduct supervised training to learn the conversion from text and desired emotion to its emotional speech, focusing on a single emotion per text-speech pair. These models only learn the correct emotional outputs without fully comprehending other emotion characteristics, which limits their capabilities of capturing the nuances between different emotions.
We propose a controllable Emo-DPO approach, which employs direct preference optimization to differentiate subtle emotional nuances between emotions through optimizing towards preferred emotions over less preferred emotional ones.
Instead of relying on traditional neural architectures used in existing emotional TTS models, we propose utilizing the emotion-aware LLM-TTS neural architecture to leverage LLMs' in-context learning and instruction-following capabilities. Comprehensive experiments confirm that our proposed method outperforms the existing baselines. 
\end{abstract}

\vspace{-0.3cm}
\section{Introduction}
\vspace{-0.1cm}
Humans produce speech that naturally varies in different emotions \cite{yasuda2023text,chen2023vector,khanam2022text,nose2007style}.
Emotional speech synthesis aims to replicate this complexity by generating human-like speech from text and the desired emotional tone, with significant advancements achieved through machine learning techniques~\cite{zhou2022speech,diatlova2023emospeech,lee2017emotional,li2024mm}.
To generate realistic emotional speech, emotional text-to-speech (TTS) models must account for various factors beyond simple text input, such as the nuanced expression of emotions through stress, intonation, rhythm, and the complex interplay of human emotional characteristics \cite{nose2007style,wang2018style}.

Current emotional TTS models predominantly rely on traditional architectures such as LSTM \cite{lei2021fine}, BLSTM~\cite{liu2024emotion}, Tacotron \cite{li2021controllable,um2020emotional,li2024mm,wang2018style}, FastSpeech~\cite{kim2021expressive,diatlova2023emospeech,lee2017emotional,li2024mm,yang2024instructtts}, VITS~\cite{zhao2023emotion}, diffusion-based model \cite{guo2023emodiff} and the flow-matching model \cite{wu2024laugh}. They neglect the integration of large language models (LLMs) to enhance speech synthesis with LLMs' in-context learning and instruction-following capabilities regarding quality, naturalness and emotional expressiveness.
In contrast, LLMs have demonstrated success in advancing speech synthesis by effectively modeling speech tokens~\cite{kimclam} and achieving high-quality synthesized voices in zero-shot scenarios~\cite{du2024cosyvoice, wang2023neural}. 
Despite this, the application of LLMs for emotion rendering in TTS models remains underexplored. This paper aims to address this gap by investigating the application of LLMs to enhance emotional speech synthesis, particularly in capturing the nuanced distinctions between different emotions.

% Reinforcement learning from human feedback (RLHF)~\cite{ouyang2022training} is core to the success of the recent powerful LLMs~\cite{achiam2023gpt,team2023gemini} and generation models~\cite{zhang2024speechalign,cideron2024musicrl}. However, due to the inherent training instability and the high computational costs associated with reinforcement learning, recent research has increasingly focused on direct alignment from preference techniques~\cite{gao2024towards}. These approaches circumvent the need for training a reward model to approximate human preferences and rely on implicit reward signals directly obtained from preference data. Among them, direct preference optimization (DPO) is a representative and highly effective technique. 

Supervised learning is predominantly used in training existing emotional TTS models, where text is paired with corresponding emotional speech, typically focusing on a single emotion per instance~\cite{cai2021emotion,diatlova2023emospeech,lee2017emotional,li2024mm}. This constrains the model's control over multiple emotions and hinders its capacity to capture subtle differences in prosody and intonation among emotions. 
To address this, we draw inspiration from reinforcement learning from human feedback (RLHF)~\cite{ouyang2022training} and direct preference optimization (DPO)~\cite{rafailov2023direct}. DPO has recently demonstrated remarkable effectiveness in distinguishing preferred signals from less preferred ones in LLMs \cite{gao2024towards,rafailov2023direct,dubey2024llama} and generative models~\cite{zhang2024speechalign,cideron2024musicrl,na2024boost,wallace2024diffusion}. RLHF, which underpins the success of modern LLMs \cite{ouyang2022training,achiam2023gpt,team2023gemini}, requires training a reward model to approximate human preferences, while DPO offers a more efficient way of optimizing preference data directly, eliminating the need for an explicit reward model and reducing the computational burden~\cite{zhang2024speechalign,cideron2024musicrl}.

\begin{figure*}[t]
\vspace{-0.6cm}
\centering
\includegraphics[width=176mm]{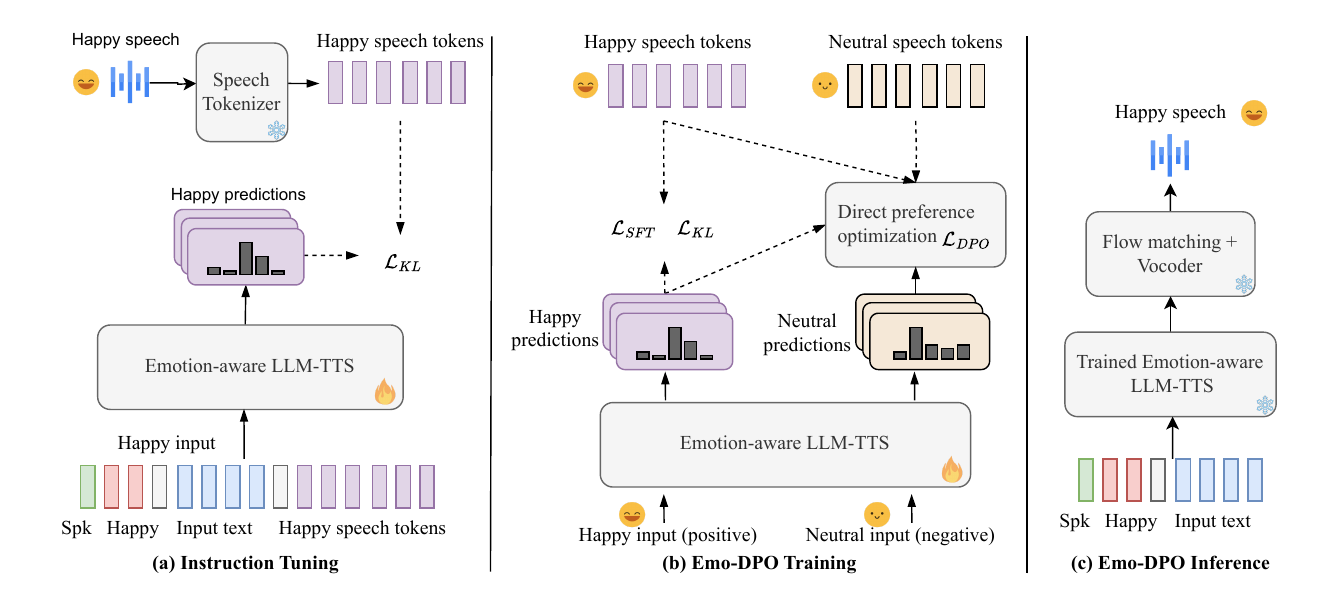}
\vspace{-0.7cm}
\caption{Overview of the proposed \textit{Emo-DPO} approach: (a) instruction tuning, (b) Emo-DPO training, and (c) the inference process.}
\label{overall}
\vspace{-0.4cm}
\end{figure*}

Motivated by the success of DPO and its role in preference alignment, we propose leveraging DPO to address the limitations of conventional emotional TTS models that control only individual emotions.
We introduce \textit{Emo-DPO}, an emotional TTS approach utilizing DPO to capture the nuanced prosodic and intonational differences between positive-negative emotion pairs, thereby enhancing emotional expressiveness in speech synthesis. Unlike traditional supervised learning methods that lack emotional preference, our \textit{Emo-DPO} fine-tunes the TTS model by aligning it with preferred emotional expression, optimizing the generation of preferred emotional outputs over less favored ones.
By incorporating both positive and negative emotional feedback, \textit{Emo-DPO} enables expressive speech synthesis go beyond single emotion modeling, and thereby better differentiating between emotions and generating more controllable and expressive emotional speech.

Our contributions of this paper include: 1) \textbf{Beyond Single Emotions}: we propose \textit{Emo-DPO}, a novel controllable emotional TTS approach that leverages direct preference optimization to differentiate subtle differences between emotions for the first time, and
2) \textbf{Emotion-aware LLM-TTS}: we investigate the integration of emotion-aware LLMs within emotional TTS neural architectures.

\section{Methodology}
\vspace{-0.1cm}
We propose an \textit{Emo-DPO} method for \textbf{emo}tional TTS through \textbf{d}irect \textbf{p}reference \textbf{o}ptimization (DPO) with an LLM-based TTS neural architecture, as illustrated in Fig~\ref{overall}.

\vspace{-0.1cm}
\subsection{\textit{Emo-DPO} Overview}
\vspace{-0.1cm}
We propose an emotional TTS approach \textit{Emo-DPO} that aims to synthesize emotional speech from text, speaker x-vector, and desired emotion inputs. Our approach combines (a) instruction tuning and (b) \textit{Emo-DPO} training with an integration of Emotion-aware LLM-TTS, optimizing the likelihood of generating a speech token sequence that corresponds to the specified emotional prompt in predefined instruction data. During inference, \textit{Emo-DPO} generates speech tokens from text, desired emotion and speaker x-vector inputs, followed by a frozen flow-matching model and a frozen vocoder to produce emotional speech (see Fig~\ref{overall} (c)). We next detail the proposed instruction tuning and \textit{Emo-DPO} training processes.

\vspace{-0.1cm}
\subsection{Instruction Tuning}
\vspace{-0.1cm}

In the first stage, we propose to perform supervised finetuning on LLM-TTS $\pi$ to benefit from LLM's instruction-following and in-context learning capabitlities using parallel emotion text-to-speech data $D_\texttt{sft}$ as shown in Fig~\ref{overall} (a). The data is formatted with the following instruction template: 
\[
d_j \in D_\texttt{sft} = E. \texttt{<endofprompt>} x_j \texttt{</s>} y_j^{+} \texttt{</s>}
\]
where $E$, $x_j$, $y_j^{+}$, $\texttt{<endofprompt>}$, and $\texttt{</s>}$ denote the emotion prompt word, such as Happy and Angry, the text token sequence, the speech token sequence that corresponds to $E$, the special token indicating the end of emotion trigger, and the separator token, respectively. The speech tokenizer extracts the speech token sequence, while the LLM-TTS model, comprising a text encoder and an LLM-based decoder, predicts the probability distribution of emotional speech tokens (eg. happy). Following~\cite{du2024cosyvoice}, we apply a label smoothing Kullback-Leibler (KL) loss to minimize the divergence between the probability distribution prediction induced by $\pi$, $P_{\pi}$ and the target (happy) distribution $P$:

\small
\[
\mathcal{L}_{KL} = KL(P_{\pi}||P) = \mathbb{E}_{d_j\sim D_\texttt{sft}} \left[ p(y^+_j | E, x_j) \log \frac{p(y^+_j | E, x_j)}{p_{\pi}(y^+_j | E, x_j)} \right]
\]
\normalsize
In this way, $\pi$ learns to generate speech token sequences that align with the specified emotional prompt in the input text, ensuring that the generated speech reflects the desired emotion as indicated by $E$.
\vspace{-0.1cm}
\subsection{Emo-Direct Preference Optimization Training}
\vspace{-0.1cm}
\textbf{Motivation:} Yet, simply conducting instruction tuning on $\pi$ may be insufficient, as the model only learns to generate the correct output without fully understanding why it is correct. To equip the model with the ability to capture subtle differences between the desired emotional speech and other emotions with the same semantic content, we turn to preference learning to further refine its performance. DPO~\cite{rafailov2023direct} provides an effective solution, allowing the model to learn directly from preference data. This ensures that the generated speech aligns more closely with the intended emotional nuances. 

\subsubsection{Beyond One Emotion - DPO Training} To construct pairwise preference data for \textit{Emo-DPO} fine-tuning (see Fig~\ref{overall} (b)), we treat $d_j$ defined above as the positive instance (eg. happy). For the negative instance, we sample from the training data other instances that share the same $x_j$ (text input) but have different emotional speech outputs (eg. neutral). Formally, the paired data $(d_j^+, d_j^-) \in D_\texttt{pref}$ is formulated as $E. \texttt{<endofprompt>} x_j \texttt{</s>} y_j^{+} \texttt{</s>}$ and $E. \texttt{<endofprompt>} x_j \texttt{</s>} y_j^{-} \texttt{</s>}$.

Denote the LLM-TTS model after the first-stage instruction tuning as $\pi_{\texttt{sft}}$. Given the pairwise dataset $D_\texttt{pref}$ and LLM-TTS $\pi$ to optimize, the DPO objective is defined as:

\begin{equation*}
\begin{aligned}
  \mathcal{L}_{\texttt{DPO}}(\pi; \pi_{\texttt{sft}}) = -\mathbb{E}_{(d_j^+, d_j^-) \sim D_\texttt{pref}} \Bigg[ 
  & \log \sigma \Big( \beta \log \frac{\pi(y^{+}_j | E, x_j)}{\pi_{\texttt{sft}}(y^{+}_j | E, x_j)} \\ 
  & - \beta \log \frac{\pi(y^{-}_j | E, x_j)}{\pi_{\texttt{sft}}(y^{-}_j | E, x_j)} \Big) \Bigg]
\end{aligned}
\end{equation*}
where $\pi$ is initialized as $\pi_{\texttt{sft}}$. $\pi(\cdot)$ refers to the conditional probability of $\pi$ generating the output sequence. $\beta$ is the hyperparameter that modulates the sharpness of $\pi$'s preference of $y_j^+$ over $y_j^-$. $\sigma$ is the sigmoid function. The DPO objective essentially maximizes the likelihood of $\pi$ generating $y_j^+$ while minimizing the likelihood of generating $y_j^-$ conditioned on $x_j$ and the emotion trigger word $E$. 

\subsubsection{Emo-DPO Training Objective} To further stabilize the training, we introduce two regularization strategies. 
One strategy is to introduce a Jensen-Shannon (JS) divergence~\cite{menendez1997jensen} manipulation to the DPO objective: 

\begin{equation*}
\begin{aligned}
&\text{(1) logits} = logratio_{\texttt{chosen}} - logratio_{\texttt{reject}} \\
                  &= \log \left( \frac{\pi(y^{+}_j | E, x_j)}{\pi_{\texttt{sft}}(y^{+}_j | E, x_j)} \right) 
                     - \log \left( \frac{\pi(y^{-}_j | E, x_j)}{\pi_{\texttt{sft}}(y^{-}_j | E, x_j)} \right) \\
&\text{(2) JSD}   = \log \left( 1 + e^{logratio_{\texttt{chosen}}} \right)
                     - \log \left( 1 + e^{logratio_{\texttt{reject}}} \right) \\
& \text{(3) logits} = \text{logits} - \text{JSD} \\
& \text{(4) } \mathcal{L}_{\texttt{DPO}}(\pi; \pi_{\texttt{sft}}) = - \mathbb{E}_{(d_j^+, d_j^-) \sim D_{\texttt{pref}}} \left[ \log \sigma \left( \beta \cdot \text{logits} \right) \right]
\end{aligned}
\end{equation*}
The above operations smooth the optimization process and prevent extreme logit differences, thus improving training stability. Additionally, they provide a more balanced and interpretable preference learning process through the bounded and symmetric nature of JS divergence. 

The other strategy is to jointly optimize the JS-regularized DPO objective, the label-smoothing KL objective defined in stage 1 of instruction tuning, and an additional SFT objective. Specifically, the total loss term is defined as:

\begin{equation*}
\mathcal{L} = \alpha \mathcal{L}_{\texttt{DPO}} + \gamma \mathcal{L}_{\texttt{KL}} + \theta \mathcal{L}_{\texttt{SFT}}
\end{equation*}
where $\mathcal{L}_{\texttt{SFT}} = -\log(\pi(y^{+}_j | E, x_j))$ while $\alpha$, $\gamma$, and $\theta$ are the hyperparameters that control the strength of each loss term. Both the label-smoothing KL loss and the SFT loss help stabilize the training by ensuring that the model remains aligned with the pre-trained LLM-TTS distribution while progressively adapting to the task-specific emotional speech generation. The JS-regularized DPO loss, on the other hand, enables the model to learn nuanced preferences from pairwise comparisons, guiding the model towards more refined and emotionally aligned outputs.

\begin{table*}
\vspace{-0.3cm}
\centering
\caption{Objective evaluation results comparison of the proposed Emo-DPO with baselines on emotion similarity, prosody similarity, intelligibility and speech emotion recognition accuracy.}
\begin{tabular}{cccc|ccccc}
\toprule
 & \textbf{Emo SIM} & \textbf{Prosody SIM} & \textbf{Intelligibility} &  & \textbf{Speech} & \textbf{Emotion} & \textbf{Recognition} &  \\
\textbf{TTS models} &  &  &  & Neutral & Angry & Happy & Sad & Surprise \\\midrule
emoepeech~\cite{diatlova2023emospeech} & 98.26 & 3.35 & 7.17 & 0.24 & 0.01 & 0.00 & 0.55 & 0.56 \\
cosyvoice~\cite{du2024cosyvoice} & 98.73 & 3.69 & 4.94 & 0.69 & 0.83 & 0.60 & 0.69 & 0.65 \\
Emo-DPO & \textbf{98.87} & \textbf{3.89} & \textbf{4.54} & \textbf{0.76} & \textbf{0.84} & \textbf{0.60} & \textbf{0.71} & \textbf{0.72} \\
\bottomrule
\end{tabular}
\label{objective}
\vspace{-0.2cm}
\end{table*}

\section{Experiments}

\subsection{Datasets and Experimental Setup}
We use English part of the ESD dataset \cite{zhou2022emotional} for experiments, with 10 speakers expressing 5 emotions: Angry, Happy, Sad, Surprise, and Neutral, with 350 utterances per speaker and emotion (about 1750 utterances and 1.2 hours per speaker). We follow official train/valid/test splits \cite{zhou2022emotional,diatlova2023emospeech}, where the validation and test sets consists of 20 and 30 utterances in 5 emotions and 10 speakers, resulting in 1000 and 1500 utterances.
We use Cosyvoice-300M-Instruct model (cosyvoice)~\cite{du2024cosyvoice} and fastspeech2 based emospeech~\cite{diatlova2023emospeech} as strong baselines,  both with publicly accessible codes. The same X-vectors for both cosyvoice and the proposed \textit{Emo-DPO} are extracted from training data for test speakers. \textit{Emo-DPO} is trained for 2 epochs with dynamic batching, followed by 3-epoch DPO training with a batch size of 8 on 4 GPUs.  
TTS-LLM, speech tokenizer, and text encoder in \textit{Emo-DPO} are initialized from cosyvoice, with the same architectures, and inference uses a pretrained flow-matching model and HifiGan vocoder~\cite{du2024cosyvoice}. Parameters $\alpha$, $\theta$ and $\gamma$ are set to 1 and other settings follow cosyvoice. For \textit{Emo-DPO} training, we create pairwise preference data with the same text by marking desired emotion audio as preferred (e.g., happy) and other emotion audio (e.g., neutral) as dis-preferred. 
\vspace{-0.3cm}
\subsection{Evaluation Metrics}
\vspace{-0.2cm}
Extensive objective and subjective evaluations are conducted to compare the proposed \textit{Emo-DPO} with baselines.

\textbf{Objective evaluations}: to assess the \textbf{intelligibility} of generated audio, we apply Whisper-Large-v3 on the audios to recognize the text and calculate the word-error-rate (WER).
\textbf{Prosody similarity (SIM)}: we use AutoPCP \cite{seamless2023} as an utterance-level estimator to quantify the prosody similarity between generated and ground-truth speech samples~\footnote{\url{https://github.com/facebookresearch/seamless_communication}} following \cite{wu2024laugh}.
\textbf{Emotion Similarity (SIM)}: we use the emotion2vec-base model~\cite{ma2023emotion2vec} to extract emotion embeddings from ground-truth and generated audio, computing cosine similarity and averaging the results across the test set for the EMO SIM score. \textbf{Speech emotion recognition} is conducted using the pretrained model~\footnote{\url{ https://huggingface.co/emotion2vec/emotion2vec_plus_large}} on the generated audios to identify the emotion categories, where Scores of 1 and 0 indicate correct and incorrect emotion identifications, respectively. Averaged scores over 1,500 test utterances are computed for each system.

\textbf{Subjective evaluations} include mean opinion score (MOS), emotion mean opinion score (Emotion MOS) and AB preference test. 20 listeners participate in all tests. \textbf{MOS} rates overall audio quality and naturalness from 1 (bad) to 5 (excellent), while \textbf{Emotion MOS} scores the similarity of emotion between the ground-truth audio and the generated speech from 1 (not at all similar) to 5 (extremely similar). In \textbf{AB preference tests}, listeners choose the better one between samples from two systems (A and B) based on quality and emotion generation.  Two AB tests are conducted: cosy vs. \textit{Emo-DPO} and emospeech vs. \textit{Emo-DPO}, each using 8 balanced emotion samples. For MOS and Emotion MOS tests, listeners are asked rate 30 samples with balanced emotions (6 samples per emotion) for cosyvoice, emospeech and \textit{Emo-DPO} models.

\vspace{-0.1cm}
\section{Results and Discussion}
\vspace{-0.1cm}
We study the effects of multiple emotion control, emotion-aware LLM-TTS integration, SFT training, DPO training and training objective design. We present objective evaluation results in Table~\ref{objective} and subjective evaluation results in Fig.~\ref{mos} and Fig~\ref{AB}. We also conduct an ablation study in Table~\ref{ablation}.

\subsection{Effectiveness of Emo-DPO training on LLM-TTS}
\begin{figure}[t]
\centering
\vspace{-0.4cm}
\includegraphics[width=73mm]{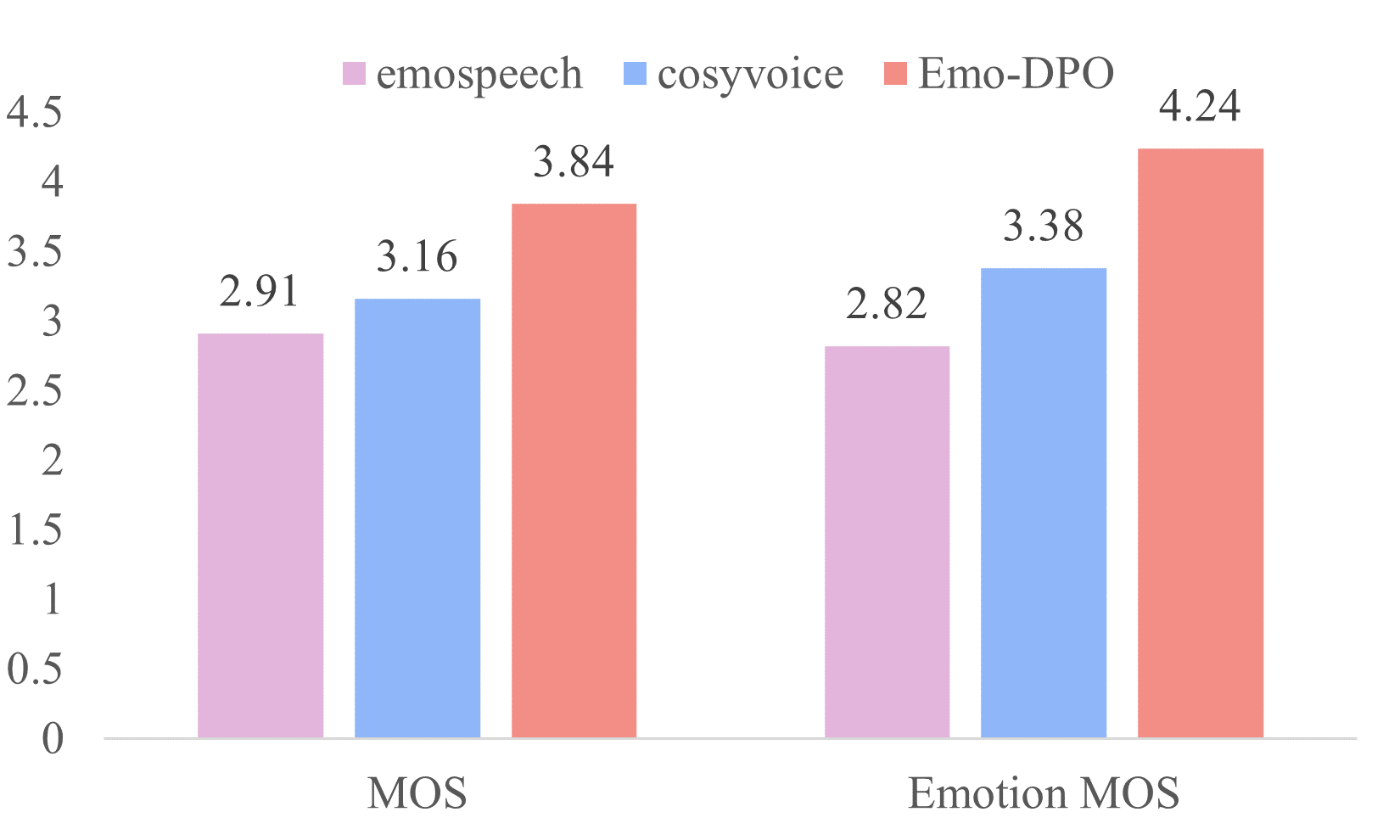}
\vspace{-0.2cm}
\caption{Comparison of subjective evaluation results for MOS and Emotion MOS tests across cosyvoice, emospeech, and the proposed \textit{Emo-DPO} models.}
\label{mos}
\vspace{-0.2cm}
\end{figure}

To assess the effectiveness of DPO training for emotional TTS, we compare baseline models (emospeech, cosyvoice) with the proposed \textit{Emo-DPO} in Table~\ref{objective}. \textit{Emo-DPO} outperforms the baselines in intelligibility, prosody similarity, and emotion similarity, demonstrating its ability to capture more subtle emotional and prosodic nuances for emotional TTS. A similar trend is seen in subjective evaluations (MOS and emotion MOS) in Fig.~\ref{mos}, showing that \textit{Emo-DPO} excels in speech quality, naturalness, and diverse emotion control. This confirms the success of DPO training in advancing emotional TTS toward more controllable, higher-quality performance. Speech emotion recognition results show that \textit{Emo-DPO} outperforms baselines, generating more controllable speech across emotions, especially for sad and surprised audios.

\begin{figure}[t]
\vspace{-0.5cm}
\centering
\includegraphics[width=89mm]{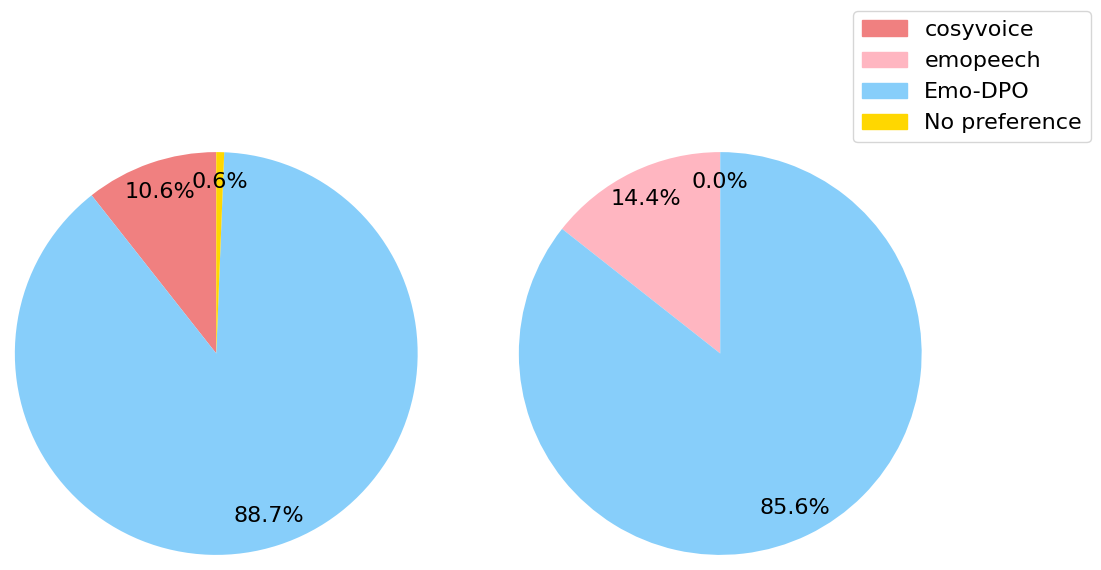}
\vspace{-0.7cm}
\caption{Comparison of subjective evaluation results from AB preference tests: 1) left: cosyvoice vs. Emo-DPO and 2) right: emospeech vs. Emo-DPO.}
\label{AB}
\vspace{-0.3cm}
\end{figure}

To facilitate a clear comparison of TTS model performance, we present AB preference test results in Fig.~\ref{AB}, showing that 85.6\% of listeners preferred the proposed \textit{Emo-DPO} over emospeech, highlighting the advantage of integrating emotion-aware LLM-TTS architecture over the traditional FastSpeech2. The superiority of the propose \textit{Emo-DPO} (88.7 \%) over cosyvoice (10.6 \%) demonstrates the enhanced ability of DPO training to capture nuanced emotional details through pairwise preference guidance. A demo page with audio samples of this work is available in the link \footnote{\url{https://xiaoxue1117.github.io/Emo-tts-dpo/}}.

\subsection{Ablation Study}
To analyze the sources of contributions, we perform an ablation study on the proposed \textit{Emo-DPO} in Table~\ref{ablation}. We observe that removing the DPO loss leads to a decline in intelligibility and prosody similarity performance, indicating that the DPO loss contributes to clearer linguistic pronunciation and better capture of diverse, time-varying prosody changes. Further removal of the SFT loss results in decreased emotional similarity, suggesting that the SFT loss helps stabilize training. 
Omitting the DPO, SFT, and KL losses leads to an overall performance drop, highlighting the effectiveness of the proposed optimization design. Additionally, removing instruction tuning and further omitting the SFT loss results in worse performance compared to the proposed model across all evaluation metrics, underscoring the importance of instruction tuning in capturing in-domain emotional characteristics.

\begin{table}
\vspace{-0.2cm}
\centering
\caption{Ablation study on the proposed Emo-DPO with different components removed w.r.t speech synthesis performances. Symbol "$-$" is removal operation.}
\begin{tabular}{lccc}
\toprule
\textbf{TTS Models} & \textbf{Emo SIM} & \textbf{Prosody SIM} & \textbf{Intelligibility} \\\midrule
Emo-DPO & \textbf{98.87} & \textbf{3.89} &  4.54
\\\midrule
- $\mathcal{L}_{\texttt{DPO}}$ & 98.87 & 3.64 &  4.94
\\
- $\mathcal{L}_{\texttt{DPO}}$ - $\mathcal{L}_{\texttt{SFT}}$ & 98.77 & 3.78 &  4.52
\\
- $\mathcal{L}_{\texttt{DPO}}$ - $\mathcal{L}_{\texttt{SFT}}$ - $\mathcal{L}_{\texttt{KL}}$ & 98.73 & 3.69 &  4.94
\\
- Instruction Tuning & 98.74 & 3.83 &  4.80\\
- Instruction Tuning - $\mathcal{L}_{\texttt{SFT}}$   & 98.80 & 3.72 &  4.55\\\bottomrule
\end{tabular}
\label{ablation}
\vspace{-0.4cm}
\end{table}

\vspace{-0.2cm}
\section{Conclusion}
\vspace{-0.2cm}
This paper presents a controllable emotional TTS approach with the integration of emotion-aware TTS-LLM architecture, opening doors for advancing emotional speech synthesis in the era of LLMs. Our proposed \textit{Emo-DPO} approach utilizes novel direct preference optimization with advanced objective designs to capture subtle emotional nuances by favoring preferred emotions over less preferred ones. Extensive experiments validate the effectiveness of Emo-DPO. 
The codes will be released upon acceptance for research community.

\footnotesize
\bibliographystyle{IEEEbib}
\bibliography{refs}

\begin{thebibliography}{10}

\bibitem{yasuda2023text}
Yusuke Yasuda and Tomoki Toda,
\newblock ``Text-to-speech synthesis based on latent variable conversion using diffusion probabilistic model and variational autoencoder,''
\newblock in {\em IEEE ICASSP}, 2023, pp. 1--5.

\bibitem{chen2023vector}
Li-Wei Chen, Shinji Watanabe, and Alexander Rudnicky,
\newblock ``A vector quantized approach for text to speech synthesis on real-world spontaneous speech,''
\newblock {\em arXiv preprint arXiv:2302.04215}, 2023.

\bibitem{khanam2022text}
Fahima Khanam, Farha~Akhter Munmun, Nadia~Afrin Ritu, Aloke~Kumar Saha, and Muhammad Firoz,
\newblock ``Text to speech synthesis: A systematic review, deep learning based architecture and future research direction,''
\newblock {\em Journal of Advances in Information Technology Vol}, vol. 13, no. 5, 2022.

\bibitem{nose2007style}
Takashi Nose, Junichi Yamagishi, Takashi Masuko, and Takao Kobayashi,
\newblock ``A style control technique for hmm-based expressive speech synthesis,''
\newblock {\em IEICE TRANSACTIONS on Information and Systems}, vol. 90, no. 9, pp. 1406--1413, 2007.

\bibitem{zhou2022speech}
Kun Zhou, Berrak Sisman, Rajib Rana, Bj{\"o}rn~W Schuller, and Haizhou Li,
\newblock ``Speech synthesis with mixed emotions,''
\newblock {\em IEEE Transactions on Affective Computing}, vol. 14, no. 4, pp. 3120--3134, 2022.

\bibitem{diatlova2023emospeech}
Daria Diatlova and Vitalii Shutov,
\newblock ``Emospeech: guiding fastspeech2 towards emotional text to speech,''
\newblock in {\em 12th Speech Synthesis Workshop (SSW) 2023}.

\bibitem{lee2017emotional}
Younggun Lee, Azam Rabiee, and Soo-Young Lee,
\newblock ``Emotional end-to-end neural speech synthesizer,''
\newblock {\em arXiv preprint arXiv:1711.05447}, 2017.

\bibitem{li2024mm}
Xiang Li, Zhi-Qi Cheng, Jun-Yan He, Xiaojiang Peng, and Alexander~G Hauptmann,
\newblock ``Mm-tts: A unified framework for multimodal, prompt-induced emotional text-to-speech synthesis,''
\newblock {\em arXiv preprint arXiv:2404.18398}, 2024.

\bibitem{wang2018style}
Yuxuan Wang, Daisy Stanton, Yu~Zhang, RJ-Skerry Ryan, Eric Battenberg, Joel Shor, Ying Xiao, Ye~Jia, Fei Ren, and Rif~A Saurous,
\newblock ``Style tokens: Unsupervised style modeling, control and transfer in end-to-end speech synthesis,''
\newblock in {\em International conference on machine learning}. PMLR, 2018, pp. 5180--5189.

\bibitem{lei2021fine}
Yi~Lei, Shan Yang, and Lei Xie,
\newblock ``Fine-grained emotion strength transfer, control and prediction for emotional speech synthesis,''
\newblock in {\em 2021 IEEE Spoken Language Technology Workshop (SLT)}. IEEE, 2021, pp. 423--430.

\bibitem{liu2024emotion}
Rui Liu, Yifan Hu, Yi~Ren, Xiang Yin, and Haizhou Li,
\newblock ``Emotion rendering for conversational speech synthesis with heterogeneous graph-based context modeling,''
\newblock in {\em Proceedings of the AAAI Conference on Artificial Intelligence}, 2024, vol.~38, pp. 18698--18706.

\bibitem{li2021controllable}
Tao Li, Shan Yang, Liumeng Xue, and Lei Xie,
\newblock ``Controllable emotion transfer for end-to-end speech synthesis,''
\newblock in {\em 2021 12th International Symposium on Chinese Spoken Language Processing (ISCSLP)}. IEEE, 2021, pp. 1--5.

\bibitem{um2020emotional}
Se-Yun Um, Sangshin Oh, Kyungguen Byun, Inseon Jang, ChungHyun Ahn, and Hong-Goo Kang,
\newblock ``Emotional speech synthesis with rich and granularized control,''
\newblock in {\em ICASSP 2020-2020 IEEE International Conference on Acoustics, Speech and Signal Processing (ICASSP)}. IEEE, 2020, pp. 7254--7258.

\bibitem{kim2021expressive}
Minchan Kim, Sung~Jun Cheon, Byoung~Jin Choi, Jong~Jin Kim, and Nam~Soo Kim,
\newblock ``Expressive text-to-speech using style tag,''
\newblock {\em arXiv preprint arXiv:2104.00436}, 2021.

\bibitem{yang2024instructtts}
Dongchao Yang, Songxiang Liu, Rongjie Huang, Chao Weng, and Helen Meng,
\newblock ``Instructtts: Modelling expressive tts in discrete latent space with natural language style prompt,''
\newblock {\em IEEE/ACM Transactions on Audio, Speech, and Language Processing}, 2024.

\bibitem{zhao2023emotion}
Wei Zhao and Zheng Yang,
\newblock ``An emotion speech synthesis method based on vits,''
\newblock {\em Applied Sciences}, vol. 13, no. 4, pp. 2225, 2023.

\bibitem{guo2023emodiff}
Yiwei Guo, Chenpeng Du, Xie Chen, and Kai Yu,
\newblock ``Emodiff: Intensity controllable emotional text-to-speech with soft-label guidance,''
\newblock in {\em ICASSP 2023-2023 IEEE International Conference on Acoustics, Speech and Signal Processing (ICASSP)}. IEEE, 2023, pp. 1--5.

\bibitem{wu2024laugh}
Haibin Wu, Xiaofei Wang, Sefik~Emre Eskimez, Manthan Thakker, Daniel Tompkins, Chung-Hsien Tsai, Canrun Li, Zhen Xiao, Sheng Zhao, Jinyu Li, et~al.,
\newblock ``Laugh now cry later: Controlling time-varying emotional states of flow-matching-based zero-shot text-to-speech,''
\newblock {\em arXiv preprint arXiv:2407.12229}, 2024.

\bibitem{kimclam}
Jaehyeon Kim, Keon Lee, Seungjun Chung, and Jaewoong Cho,
\newblock ``Clam-tts: Improving neural codec language model for zero-shot text-to-speech,''
\newblock in {\em The Twelfth International Conference on Learning Representations}.

\bibitem{du2024cosyvoice}
Zhihao Du, Qian Chen, Shiliang Zhang, Kai Hu, Heng Lu, Yexin Yang, Hangrui Hu, Siqi Zheng, Yue Gu, Ziyang Ma, et~al.,
\newblock ``Cosyvoice: A scalable multilingual zero-shot text-to-speech synthesizer based on supervised semantic tokens,''
\newblock {\em arXiv preprint arXiv:2407.05407}, 2024.

\bibitem{wang2023neural}
Chengyi Wang, Sanyuan Chen, Yu~Wu, Ziqiang Zhang, Long Zhou, Shujie Liu, Zhuo Chen, Yanqing Liu, Huaming Wang, Jinyu Li, et~al.,
\newblock ``Neural codec language models are zero-shot text to speech synthesizers,''
\newblock {\em arXiv preprint arXiv:2301.02111}, 2023.

\bibitem{cai2021emotion}
Xiong Cai, Dongyang Dai, Zhiyong Wu, Xiang Li, Jingbei Li, and Helen Meng,
\newblock ``Emotion controllable speech synthesis using emotion-unlabeled dataset with the assistance of cross-domain speech emotion recognition,''
\newblock in {\em ICASSP 2021-2021 IEEE International Conference on Acoustics, Speech and Signal Processing (ICASSP)}. IEEE, 2021, pp. 5734--5738.

\bibitem{ouyang2022training}
Long Ouyang, Jeffrey Wu, Xu~Jiang, Diogo Almeida, Carroll Wainwright, Pamela Mishkin, et~al.,
\newblock ``Training language models to follow instructions with human feedback,''
\newblock in {\em Advances in Neural Information Processing Systems}, S.~Koyejo, S.~Mohamed, A.~Agarwal, D.~Belgrave, K.~Cho, and A.~Oh, Eds. 2022, vol.~35, pp. 27730--27744, Curran Associates, Inc.

\bibitem{rafailov2023direct}
Rafael Rafailov, Archit Sharma, Eric Mitchell, Christopher~D Manning, Stefano Ermon, and Chelsea Finn,
\newblock ``Direct preference optimization: Your language model is secretly a reward model,''
\newblock in {\em Thirty-seventh Conference on Neural Information Processing Systems}, 2023.

\bibitem{gao2024towards}
Bofei Gao, Feifan Song, Yibo Miao, Zefan Cai, Zhe Yang, Liang Chen, Helan Hu, Runxin Xu, Qingxiu Dong, Ce~Zheng, Wen Xiao, Ge~Zhang, Daoguang Zan, Keming Lu, Bowen Yu, Dayiheng Liu, Zeyu Cui, Jian Yang, Lei Sha, Houfeng Wang, Zhifang Sui, Peiyi Wang, Tianyu Liu, and Baobao Chang,
\newblock ``Towards a unified view of preference learning for large language models: A survey,''
\newblock {\em arXiv preprint arXiv: 2409.02795}, 2024.

\bibitem{dubey2024llama}
Abhimanyu Dubey, Abhinav Jauhri, Abhinav Pandey, Abhishek Kadian, Ahmad Al-Dahle, Aiesha Letman, Akhil Mathur, Alan Schelten, Amy Yang, Angela Fan, et~al.,
\newblock ``The llama 3 herd of models,''
\newblock {\em arXiv preprint arXiv:2407.21783}, 2024.

\bibitem{zhang2024speechalign}
Dong Zhang, Zhaowei Li, Shimin Li, Xin Zhang, Pengyu Wang, Yaqian Zhou, and Xipeng Qiu,
\newblock ``Speechalign: Aligning speech generation to human preferences,''
\newblock {\em arXiv preprint arXiv:2404.05600}, 2024.

\bibitem{cideron2024musicrl}
Geoffrey Cideron, Sertan Girgin, Mauro Verzetti, Damien Vincent, Matej Kastelic, Zal{\'a}n Borsos, Brian McWilliams, Victor Ungureanu, Olivier Bachem, Olivier Pietquin, et~al.,
\newblock ``Musicrl: Aligning music generation to human preferences,''
\newblock {\em arXiv preprint arXiv:2402.04229}, 2024.

\bibitem{na2024boost}
Sanghyeon Na, Yonggyu Kim, and Hyunjoon Lee,
\newblock ``Boost your own human image generation model via direct preference optimization with ai feedback,''
\newblock {\em arXiv preprint arXiv:2405.20216}, 2024.

\bibitem{wallace2024diffusion}
Bram Wallace, Meihua Dang, Rafael Rafailov, Linqi Zhou, Aaron Lou, Senthil Purushwalkam, Stefano Ermon, Caiming Xiong, Shafiq Joty, and Nikhil Naik,
\newblock ``Diffusion model alignment using direct preference optimization,''
\newblock in {\em Proceedings of the IEEE/CVF Conference on Computer Vision and Pattern Recognition}, 2024, pp. 8228--8238.

\bibitem{achiam2023gpt}
Josh Achiam, Steven Adler, Sandhini Agarwal, Lama Ahmad, Ilge Akkaya, Florencia~Leoni Aleman, Diogo Almeida, Janko Altenschmidt, Sam Altman, Shyamal Anadkat, et~al.,
\newblock ``{GPT}-4 technical report,''
\newblock {\em arXiv preprint arXiv:2303.08774}, 2023.

\bibitem{team2023gemini}
Gemini~Team Google, Rohan Anil, Sebastian Borgeaud, Yonghui Wu, Jean-Baptiste Alayrac, Jiahui Yu, Radu Soricut, Johan Schalkwyk, Andrew~M Dai, Anja Hauth, et~al.,
\newblock ``Gemini: a family of highly capable multimodal models,''
\newblock {\em arXiv preprint arXiv:2312.11805}, 2023.

\bibitem{menendez1997jensen}
Mar{\'\i}a~Luisa Men{\'e}ndez, JA~Pardo, L~Pardo, and MC~Pardo,
\newblock ``The jensen-shannon divergence,''
\newblock {\em Journal of the Franklin Institute}, vol. 334, no. 2, pp. 307--318, 1997.

\bibitem{zhou2022emotional}
Kun Zhou, Berrak Sisman, Rui Liu, and Haizhou Li,
\newblock ``Emotional voice conversion: Theory, databases and esd,''
\newblock {\em Speech Communication}, vol. 137, pp. 1--18, 2022.

\bibitem{seamless2023}
Lo{\"\i}c Barrault, Yu-An Chung, Mariano~Coria Meglioli, David Dale, Ning Dong, Mark Duppenthaler, Paul-Ambroise Duquenne, Brian Ellis, Hady Elsahar, Justin Haaheim, et~al.,
\newblock ``Seamless: Multilingual expressive and streaming speech translation,''
\newblock {\em arXiv preprint arXiv:2312.05187}, 2023.

\bibitem{ma2023emotion2vec}
Ziyang Ma, Zhisheng Zheng, Jiaxin Ye, Jinchao Li, Zhifu Gao, Shiliang Zhang, and Xie Chen,
\newblock ``emotion2vec: Self-supervised pre-training for speech emotion representation,''
\newblock {\em arXiv preprint arXiv:2312.15185}, 2023.

\end{thebibliography}
\end{document}